\documentclass[aps, twocolumn, showpacs]{revtex4-1}
\usepackage{amsmath}
\begin{document}
\title{Stochasticity, topology, and spin}
\author{S. C. Tiwari \\
Department of Physics, Institute of Science, Banaras Hindu University, Varanasi 221005, and \\ Institute of Natural Philosophy \\
Varanasi India\\}
\begin{abstract}
Complex Schroedinger equation is transformed to spinor or coupled scalar field equations replacing the imaginary unit $i$ by
a matrix $\begin{bmatrix} 0 & 1 \\-1 & 0 \end{bmatrix}$. New perspecive on stochasic approach is developed with spin as topological
invariant and mass having stochastic origin in this Spinor Random Field formalism.
\end{abstract}
\maketitle
\section{\bf Introduction}

Recent advances in technology have led to the emergence of quantum information science and technology, simultaneously the foundations of 
quantum mechanics have also seen intense activity and past thought experiments are getting realized in the laboratory. To put them in perspective
Ludwig's description of the development of quantum mechanics via two routes, namely from point mechanics to Heisenberg-Jordan-Born
matrix mechanics, and wave field to Schroedinger wave mechanics is quite insightful \cite{1}. Formal equivalence of the two, and von Neumann's 
axiomatization \cite{2} led to the language of operator calculus and Hilbert state space description of the quantum system. Since the inception
of quantum mechanics the physical interpretation has been controversial and nonunique in the light of shades of diverse approaches \cite{3}. 
Obviously for a fundamental theory this sort of affairs is not good \cite{4}. Is counter-intuitiveness an intrinsic attribute at the microscopic
level? We believe that weirdness and mystery are unphysical to understand non-classical nature of a physical system \cite{4}. 

Unfortunately even for a century old concept of spin one usually finds incomprehensible picture in the literature. Spinors as geometric objects by Cartan  
appeared in 1913 \cite{5}; in physics the historical development and the enigma of the spin are reviewed in \cite{6, 7, 8}. Yet, the physical 
origin of spin seems obscure. Does Schroedinger equation represent spinless particle? Could one justify connection between zitterbewegung
and spin? Could one explain spin as angular momentum carried by some kind of a field? Though these questions have been discussed in the literature
cited in the reviews, most of them are considered as tentative models/explanations. We propose to discuss this question in a new approach in which
the Schroedinger equation is replaced by coupled equations for two real fields; and in view of real fields fluid dynamical as well as
stochastic approach become natural. Note that in the major literature on de Broglie-Bohm theory 
and stochastic quantum mechanics it becomes necessary to introduce two real scalars in the complex Schroedinger wave function
\begin{equation}
\Psi = \sqrt{\rho} e^{\frac{i S}{\hbar}} 
\end{equation}
and decompose Schroedinger wave equation into two components separating real and imaginary parts. Here $\rho$ is the probability density. 
We develop a new approach with two scalars and coupled wave equations, and show that it is
vastly general than the standard Schroedinger equation. We call our approach as 
spinor-random-field (SRF) theory for the reasons that we explain in the following.

The appearance of the imaginary unit $\sqrt{-1} =i$ in the Schroedinger equation makes it ``the mysterious wave equation'' \cite{3}. Formal
similarity between the free particle Schroedinger equation and the diffusion equation led to the Schroedinger-Furth approach. Analytic continuation
in time connects the diffusion equation and Schroedinger equation, and Feynman integral and Wiener integral. We have argued that the mass
parameter and imaginary unit $i$ deserve to be analyzed afresh \cite{9}. In fact, analytic continuation in mass has also been used \cite{10}.
Physical basis for stochastic quantum mechanics \cite{11} and brief review on mathematical aspects on Kolmogorov-Ito theory and Euclidean
quantum field theory \cite{12} are referred for details. Sustained pioneering work by de la Pena and his collaborators over past several decades
on the stochastic electrodynamics (SED) using zero-point radiation field (ZPF), in spite of great promise as an alternative to the standard 
Copenhagen quantum mechanics, continues to remain only as a viable program in need of substantial further development \cite{13}. A notable
recent advancement is in the context of the origin of spin in SED \cite{14}. On the other hand, we have explored topological origin of spin \cite{15}.
In SRF, it is possible to interpret the wave equations stochastically, hence the term ``randon field''. SRF has a potential to make
break-through progress in SED pursued in \cite{13, 14, 16}.

In the next section we derive SRF equations, and show correspondence with Schroedinger wave mechanics. The new nontrivial characteristics
of SRF theory are also pointed out. In section 3 special case of SRF equations is discussed having implications on zitterbewegung, topology and spin. 
Concluding remarks constitute the last section.

\section{\bf Spinor-Random-Field Equations: Field Interpretation}

Originally the wave equation for the mechanical field scalar was obtained by Schroedinger in his fourth communication, see page 151 in \cite{1}
that we write here
\begin{equation}
(-\frac{\hbar^2}{2m} \nabla^2 ~ + V) \Psi =\pm i \hbar \frac{\partial \Psi}{\partial t} 
\end{equation}
Here $V$ is time-independent and real. Though expression (1) substituted in Eq.(2), and separation of real and imaginary parts to 
obtain equivalent two equations for various interpretations, e. g. hydrodynamical, de Broglie-Bohm theory, and stochastic interpretation
is a justified approach, one must realize that orthodox Copenhagen interpretation and operator-observable and Hilbert space formalism
are inappropriate for them. It would seem that these are closer to original scalar field envisaged by Schroedinger \cite{1}. In a
somewhat radically different approach \cite{9} it was argued that mass has stochastic origin. Recently arbitrary choice of the free parameter
in Schroedinger-like equations has been criticized, and specific choice of Planck constant in (2) is shown to arise in SED \cite{16}.

Our main concern is the enigma of $i$ in Eq.(2). Is it possible to eliminate it? If $i$ is not present in the wave equations, stochastic
interpretation becomes more natural. A remarkably new method is found utilizing a matrix $C$ introduced in
 \cite{5}
\begin{equation}
 C =\begin{bmatrix} 0 & 1 \\ -1 & 0 \end{bmatrix}
\end{equation}
having the interesting properties
\begin{equation}
C^T = -C, ~ C C^T =1,~ C^2 = - 1 
\end{equation}
Now the last property of $C$ could be viewed as equivalent to $i^2 =-1$, and we may seek a replacement $i \rightarrow C$ in Eq.(2).
The wave function must be replaced by two components
\begin{equation}
 \Psi ~ \rightarrow ~ \begin{bmatrix} \eta \\ \chi \end{bmatrix} 
\end{equation}
Thus we arrive at new wave equations replacing the Schroedinger equation
\begin{equation}
 (-\frac{\hbar^2}{2m} \nabla^2 ~ + V) \eta = \hbar \frac{\partial \chi}{\partial t} 
\end{equation}
\begin{equation}
 (-\frac{\hbar^2}{2m} \nabla^2 ~ + V) \chi = -\hbar \frac{\partial \eta}{\partial t}
\end{equation}
Eqs. (6) and (7) in spinor form read
\begin{equation}
 (-\frac{\hbar^2}{2m} \nabla^2 ~ + V) I \begin{bmatrix} \eta \\ \chi \end{bmatrix} = \hbar \frac{\partial }{\partial t} C
 \begin{bmatrix} \eta \\ \chi \end{bmatrix} 
\end{equation}
Equations (6) to (8) represent the desired SRF equations.

Recall that Schroedinger actually obtained fourth-order equation, and drew analogy with the vibrating plate in elasticity theory \cite{1}. 
It can be easily checked that SRF equations transform to uncoupled Schroedinger's fourth-order equation. Remarkably the standard Schroedinger
wave equation (2) becomes merely a trivial special case of SRF equations: assume
\begin{equation}
 \eta = a \chi
\end{equation}
then the consistency between Eqs. (6) and (7) demands $a = \pm i$. Obviously one is free to restrict oneself to the standard quantum mechanics,
however our motivation is to explore SRF theory beyond this.

Let us first consider the usual field interpretation; following the standard prescription we derive a current continuity equation. 
Multiplying Eq.(6) by $\chi$ and Eq.(7) by $\eta$ and subtracting the resulting equations we obtain
\begin{equation}
{\bf \nabla}.\tilde{\bf J} +\frac{\partial \tilde{\rho}}{\partial t} =0 
\end{equation}
\begin{equation}
\tilde{\rho} = \frac{(\eta^2 +\chi^2)}{2} 
\end{equation}
\begin{equation}
\tilde{\bf J} = \frac{\hbar}{2 m} ( \chi {\bf \nabla} \eta - \eta {\bf \nabla} \chi)
\end{equation}
From expression (12) one may introduce a current velocity field
\begin{equation}
\tilde{\bf v} = \frac{\tilde{\bf J}}{\tilde{\rho}} 
\end{equation}
Since we do not have a complex Schroedinger wavefunction here, there is a freedom to seek (non-random) field interpretation: $\tilde{\rho}$
may be interpreted as energy density and $\tilde{\bf J}$ as momentum density of the fields. As usual, one may define angular momentum
density to be ${\bf r} \times \tilde{\bf J}$.

Note that Eq.(8) is a spinor field equation with real components of the spinor. In Cartan theory of spinors in 3-dimensional pseudo-Euclidean
space \cite{5} one indeed gets spinors with real components. Taking the limit $m \rightarrow 0$ following \cite{9}, and introducing
second-order time derivative for relativistic invariance it is easily verified that each component of the spinor satisfies d'Alember
wave equation. We suggest that all this discussion on the field interpretation of SRF hints at the presence of spin in the Schroedinger
equation perhaps hidden in the imaginary unit $i$. We elaborate on this in the next section.

\section{\bf Stochastic SRF and spin}

Quantum mechanics has proved immensely successful to understand microscopic phenomena. It would be quite logical to retain essence of 
quantum mechanics in SRF. Undoubtedly it is the probabilistic element. It is instructive to recall that in Schroedinger theory
the probability density $\rho = \Psi^* \Psi$ and probability current density
\begin{equation}
{\bf J} = \frac{\hbar}{2m i} (\Psi^* {\bf \nabla} \Psi - {\bf \nabla}\Psi^* \Psi) 
\end{equation}
satisfy the continuity equation
\begin{equation}
{\bf \nabla}. {\bf J} + \frac{\partial \rho}{\partial t} =0
\end{equation}
In analogy to this, instead of energy density, $\tilde{\rho}$ may be interpreted as probability density. Stochastic process of some kind
becomes natural. However, in contrast to the existing literature on stochastic approaches \cite{11,13,14} the new element in the
present work is the inevitable presence of spin. What does this mean?

Reviews and history of spin \cite{6,7,8} show that 1) at the abstract level, a particle is characterized by two Casimir invariants 
of the Poincare group, and for a particle with non-zero rest mass Wigner's little group is SU(2), 2) the spin operator 
${\bf S} =\frac{1}{2} \hbar {\bf \sigma}$ satisfies su(2) Lie algebra, and spin projection, let us say along z-axis, is $\pm \hbar/2$
given by the eigenvalues of $S_z$ with eigenvectors $\begin{bmatrix} 1 \\ 0 \end{bmatrix}$ and $\begin{bmatrix} 0 \\ 1 \end{bmatrix}$
for a pure state, 3) in general, the polarization vector with expectation values of all the three components of ${\bf S}$ defines
the particle state, and 4) for a zero mass particle spin is parallel or anti-parallel to the velocity, and this statement is
relativistically invariant as shown by Wigner; if there is any internal motion to explain spin it must be perpendicular to the velocity.

A remarkable, though unconventional, inference was drawn by Gurtler and Hestenes \cite{17} that ``Schroedinger theory is identical to Pauli
theory when the electron is an eigenstate of the spin''. Then there is the ``Sommerfeld puzzle'': the Sommerfeld semiclassical quantization
for relativistic hydrogen atom gives exactly the energy levels that were later obtained in Dirac's relativistic equation for the Coulomb
problem. Biedenharn \cite{18} shows that the implicit role of spin is responsible for this coincidence. SRF is developed in the present work
replacing $i$ in the Schroedinger equation by the matrix $C$, therefore, taken together with the past ideas \cite{17, 18} it is reasonable
to argue that imprint of spin is implicit in the Schroedinger equation. As an important consequence, the physical
origin of spin acquires a radically new perspective. It necessitates re-examination of the roles of SED \cite{14} and 
topology \cite{15, 19} in understanding spin.

Spin has been related with a topological invariant of half-quantized vortex in a recent geometric model of the electron \cite{15, 19}. For 
proton-spin de Rham period for a closed 3-form over a 3-cycle is identified to be a topological invariant \cite{20}.
Earlier, orbifold quantization was suggested for the photon spin \cite{21}. The attempt to relate these abstract mathematical constructs with
physical observables has also been made. Wilson lines may serve as probes of topology in proton scattering experiments \cite{20}.
Polarization-sensitive interference in an interesting experiment \cite{22} seems to support topological photon with spin determining particle
nature of the photon. Unlike proton that has extended internal constituent structure, and photon that is massless, the electron is believed
to be a point particle with non-zero mass. Obviously electron spin requires a different approach: both stochastic approach \cite{14} and
geometry/topology are important. In fact, Thomas work, nicely discussed by Tomonaga \cite{6}, has a profound observation on the parallel
transport holonomy for spin of the electron highlighted recently \cite{23}. Exact half-quantized vortex model of spin \cite{15, 19} seems to be applicable
for massless particle; it is the mass term that requires attention for stochastic explanation.

Brief review in recent article \cite{14} notes the role of ZPF and zitterbewegung, and approximate spin value acquired by the electron in
previous works on SED. The main point in section 3 of their paper \cite{14} seems to be that intrinsic spin of electron emerges from 
the coupling of the electron (particle) to the separate polarized field modes of ZPF: it is a sort of angular momentum transfer from the ZPF.
The present work suggests that hidden thermostat of de Broglie \cite{24} or ZPF in SED give mass to the electron that intrinsically is massless and has 
spin as a topological invariant. In this alternative scenario the stochastic process involves scattering of a particle moving with light velocity 
in-between the interactions, and the average spin value need not be exactly $\pm \hbar/2$, perhaps the results of \cite{25} are more appropriate.

Above idea needs to be developed for getting concrete results; however we make two remarks for further exploration. First, let us have a naive picture.
SRF equations (6) and (7) with $V=0$ show that for specific assumption
\begin{equation}
\hbar \frac{\partial \chi}{\partial t} = E \eta 
\end{equation}
SRF equations are un-coupled and transform to
\begin{equation}
-\frac{\hbar^2}{2m} \nabla^2 \eta = E \eta 
\end{equation}
\begin{equation}
\nabla^2 \chi = \frac{2 m}{E} \frac{\partial^2 \chi}{\partial t^2} 
\end{equation}
Here $E$ is a constant. The wave equation (18) assumes Lorentz covariant form
\begin{equation}
\nabla^2 \chi =\frac{1}{c^2} \frac{\partial^2 \chi}{\partial t^2} 
\end{equation}
setting
\begin{equation}
E = 2 m c^2 
\end{equation}
In this case Eq.(17) becomes
\begin{equation}
\nabla^2 \eta = - k^2 \eta 
\end{equation}
where
\begin{equation}
k = \frac{2 \pi}{\lambda}, ~ \lambda =\frac{h}{2 m c} 
\end{equation}

To interpret Eqs. (19) and (20) an illuminating, though speculative, discussion utilizing the concept of zitterbewegung in the 
stochastic interpretation \cite{26} seems interesting. Note that here we have two fields $\chi$ and $\eta$, therefore the applicability
of the stationary Schroedinger-like equation and topological defect for relativistic field become rather natural. Obviously, the traditional
stochastic approach has to be modified: a tentative proposition is given here for this purpose.

Nelson \cite{11} and Streater \cite{12} discuss various kinds of stochastic processes. Nelson has based his theory on physical arguments, and
assumes Newton's second law for a point mass particle with additional Wiener or random force. The particle has fluctuating position
due to its interaction with a hidden thermostat resulting into a Markovian process in the coordinate space. To use Newton's law one has
to define mean acceleration, towards this aim he uses Ornstein-Uhlenbeck theory. To obtain stochastic differential equation 
(like Langevin equation) for a random variable $x(t)$, where $t \in R^+$ is interpreted as time, mean forward and backward derivatives are 
respectively defines as
\begin{equation}
 D_f x(t) = Lt_{\Delta t \rightarrow 0^+} ~ \frac{< x(t+\Delta t) - x(t)>}{\Delta t}
\end{equation}
\begin{equation}
 D_b x(t) = Lt_{\Delta t \rightarrow 0^+} ~ \frac{< x(t) - x(t - \Delta t)>}{\Delta t}
\end{equation}
If $x(t)$ is differentiable $D_f =D_b =\frac{d}{dt}$. The kinematics of the Markovian process is given by the forward and backward Fokker-Planck
equations for the probability density $\hat{\rho}({\bf x},t)$ of ${\bf x}(t)$
\begin{equation}
 \frac{\partial \hat{\rho}}{\partial t} = - {\bf \nabla}.{\bf b}_f + \nu \nabla^2 \hat{\rho}
\end{equation}
\begin{equation}
 \frac{\partial \hat{\rho}}{\partial t} = - {\bf \nabla}.{\bf b}_b - \nu \nabla^2 \hat{\rho}
\end{equation}
Here $\nu$ is the diffusion constant, and forward (backward) velocity ${\bf b}_f ~({\bf b}_b)$ is a vector valued function on space-time.

Nelson's theory is a classical probabilistic theory in which Schroedinger equation is obtained introducing the complex wavefunction of the form (1),
and assuming diffusion coeifficient $\nu =\frac{\hbar}{2 m}$. There are two velocity vectors, the current velocity
\begin{equation}
 {\bf v} = \frac{1}{2} ({\bf b}_f +{\bf b}_b)
\end{equation}
and osmotic velocity
\begin{equation}
 {\bf u} = \frac{1}{2} ({\bf b}_f -{\bf b}_b)
\end{equation}

Preceding discussion on SRF offers new physical insights: the particle travels at the speed of light, mass is attributed to the influence
of the random field (hidden thermostat or ZPF), and spin originates as a topological defect, e. g. half-quantized vortex. The stochastic
process is some kind of continued scattering with the random field that changes the direction of the propagation of the defect giving
rise to a mean trajectory with average velocity less than the velocity of the light. Half-quantized vortex originates in the transverse internal
motion where the length scale is that of the Compton wavelength, $\lambda_c$. The mean distance traveled with velocity of light is
much smaller than $\lambda_c$. We suggest that time $\tau = \frac{e^2}{m c^3}$ corresponding to electron charge radius determines this distance.
Topology and geometric model of electron \cite{15, 19} are naturally incorporated in this picture with a significant role of schotasticity.

The occurrence of time scale $\frac{2}{3} \tau$, and Compton wavelength in the stochastic theory based on ZPF \cite{25, 26} shows that the 
tentative proposition suggested here has a potential to provide impetus to SED approach. The limit $\Delta t \rightarrow 0^+$ in the 
definitions (23) and (24) has to be changed as particle travels at speed $c$; may be it has to be replaced by $\tau$. Instead of mass
entering in the diffusion coefficient $\nu$, the mass itself has to be treated as a random variable. One possibility is that instead of
Newton's second law with Lorentz force in \cite{26}, one may use random potential and seek average momentum: here the pure gauge field 
of Aharonov-Bohm effect with additional random variable multiplying it could be explored.

\section{\bf Conclusion}

A new framework replacing Schroedinger equation by spinor field equation or coupled scalar field equations is developed. It is
proposed that mass has stochastic origin, and spin is a topological invariant. The importance of random fields for observed mean 
trajectory of electron has been discussed that has implications for further progress in SED approach.

The use of imaginary time and analytic continuation in quantum field theory beginning with Dyson and Schwinger has been mathematically developed
as Euclidean quantum field theory \cite{12}. Our approach may result into a new perspective in this field.

\end{document}